\documentstyle[12pt]{article}  

\input psfig

\begin{document}
{\bf   To appear in Chaos Solitons \& Fractals}\\ {\em  Special Issue on
"Chaos and Quantum Transport in Mesoscopic Cosmos", editor: K. Nakamura
(1997)}  
\vspace*{2.5 cm} 
\begin{center} 
{\bf \Large Diffraction in the semiclassical description  
of mesoscopic devices} 
\end{center} 
%\vspace{10 mm} 
\begin{center} 
{\bf G. VATTAY, J. CSERTI, G. PALLA and G. SZ\'ALKA} 
\end{center} 
%\vspace{5 mm} 
\begin{center} 
{\rm \small Institute for Solid State Physics, E\"otv\"os University, 
M\'uzeum krt. 6-8, H-1088 Budapest, Hungary}

\end{center} 
 
%\vspace{10 mm} 
\vspace{15mm} 
 
\baselineskip 7.5mm 
\parbox[t]{160mm}{\baselineskip 7.5mm 
{\bf Abstract} -In pseudo integrable systems diffractive  
scattering caused by wedges and impurities can be described within the  
framework of Geometric Theory of Diffraction (GDT) in a way similar  
to the one used in the Periodic Orbit Theory of Diffraction (POTD). 
We derive formulas expressing the reflection and transition 
matrix elements for one and many diffractive points and apply it for  
impurity and wedge diffraction.  
Diffraction can cause backscattering in situations, where usual 
semiclassical backscattering is absent causing an  erodation 
of ideal conductance steps. The length of diffractive periodic orbits  
and diffractive loops can be detected in the power spectrum of the 
reflection matrix elements. The tail of the power spectrum 
shows $\sim 1/l^{1/2}$ decay due to impurity scattering and 
$\sim 1/l^{3/2}$ 
decay due to wedge scattering. We think this is a universal sign 
of the presence of diffractive scattering in pseudo integrable  
waveguides.} 
 
\vspace{36pt}

In recent years, semiclassical methods became very popular in 
describing devices operating in the mesoscopic regime.  There are two 
entirely different sets of theoretical tools which have been used in a 
wide range of applications. One of them is a WKB based short wavelength 
description \cite{Gut,Voros} where classical trajectories, chaos, 
regularity and analytic properties of the potential play a major role. 
The other is based on  random matrix models or on averaging 
over random Gaussian potentials \cite{Efetov}.  The first approach is 
designed to describe clean systems where the potential depends smoothly 
on coordinates and parameters, while the second assumes a system 
densely packed with impurities causing wild fluctuations of the 
potential on all length scales. 
  
Fortunately, advances in manufacturing and material design reduced the 
average number of impurities exponentially since the eighties and this 
trend is expected to continue in the future.  Accordingly, a realistic 
semiclassical theory should be capable to treat systems with low number 
of impurities and not only completely clean or completely dirty ones. 
Another demand is that human designed structures, 
straight sides and eventual wedges should be a natural part of the 
description. These systems, neither chaotic nor regular, form the 
intermediate class of pseudo integrable systems. In the 
present paper we will demonstrate that diffractive scattering caused by 
wedges and impurities can be described within the framework of Geometric 
Theory of Diffraction \cite{Buslaev,Keller1,Keller2,Keller3,Keller4} (GDT) in a way similar to 
the one used in the Periodic Orbit Theory of Diffraction \cite{Vattay1,
Vattay2,Vattay3} 
(POTD).  The POTD is an extension of Gutzwiller's periodic orbit theory 
for non-smooth systems with diffractive scattering which has been used 
successfully during the last three years. 
  
\section{Examples of pseudo integrable mesoscopic devices}  
  
The simplest mesoscopic device is a two dimensional wave guide strip of 
electron gas, formed on a GaAs heterostructure \cite{Beenakker}, 
connecting two electron 
reservoirs (Fig. 1a). 
\begin{figure}
\centerline{\strut\psfig{figure=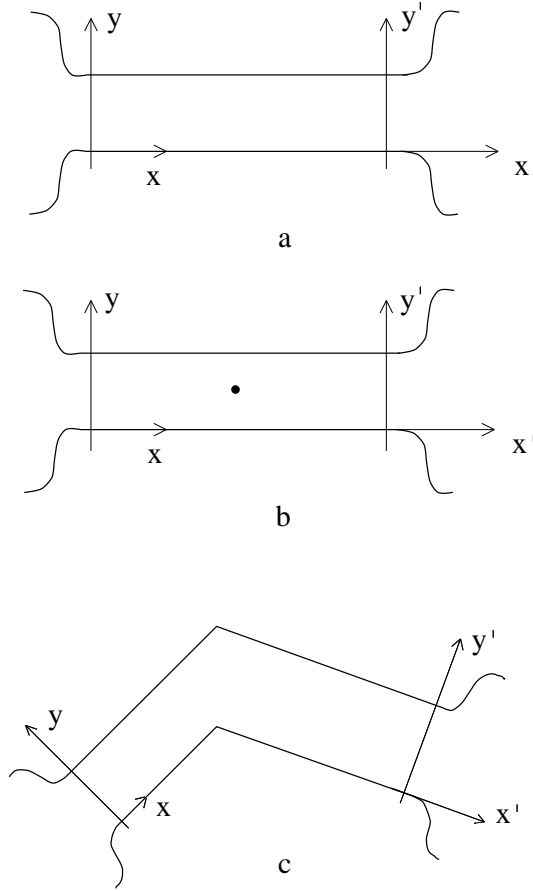,width=70mm}}
\caption{\small Electron reservoirs (Ohmic contacts) connected trough 
strip waveguides. {\bf a,} Ideal strip waveguide with parallel walls.
{\bf b,} Ideal strip waveguide with parallel walls and one pointlike 
impurity. {\bf c,} Broken-strip waveguide.} 
\end{figure}
This system can be considered also as a  paradigm 
of a more sophisticated two or three dimensional clean quantum wire 
\cite{Zarand} 
made of other materials, for example carbon based structures with 
delocalized electrons. Here we will consider two slightly modified 
versions of this simple quantum wire, the quantum wire with one point 
like impurity (see Fig. 1b) \cite{csoport1}, the broken-strip waveguide 
(see Fig. 1c) \cite{csoport2}. 
Our motivation in selecting  these systems for investigations is that 
any long wire with straight sides, many wedges and  impurities can be 
built up as a sequence of these building blocks. For simplicity, the 
walls of the wave guides are assumed to be completely hard and the wave 
function vanishes on them. The  potential is assumed to be zero inside 
the waveguide.

\section{The Landauer Formula}  
  
The conductance of the non-degenerate electron gas in a waveguide considered 
here is given by the Landauer formula \cite{Landauer,BarSto}. 
According to this theory, incoming and outgoing wave functions in the leads 
far from impurities or wedges can be decomposed into incoming and 
outgoing quantum modes
\begin{eqnarray}  
\psi_n(x,y)&=&\Phi_n(y)e^{ik_nx},\nonumber \\  
\Phi_n(y)&=&\sqrt{\frac{2}{W}}\sin(\pi n y/W), 
\end{eqnarray}    
where $k_n=\sqrt{2mE_F-(\hbar\pi n/W)^2}/\hbar$ is the wave number of 
the propagating planewave, $W$ is the width of the lead and $E_F$ is 
the Fermi energy.  If the Fermi energy is less than 
$E_n=\frac{1}{2m}(\hbar\pi n/W)^2$ the 
wavenumber $k_n$ becomes imaginary and the $n^{th}$ channel becomes closed 
preventing wave propagation.

The Landauer formula for the conductance $G(E_F)$ at Fermi
energy $E_F$ is then
\begin{equation}  
G(E_F)=\frac{2e^2}{h}\sum_{m,n} |t_{mn}|^2, 
\end{equation}  
where $t_{nm}$ is the transition probability amplitude from the 
incoming cannel $n$ on the entrance side to the outgoing channel $m$ 
on the exit side.  In case of infinitely long leads summation goes for 
open channels of both sides only.  The transition probability is given by
the projection of the Green function over the transverse
wave functions $\Phi_n(y)$ on the entrance lead for the incoming modes
and $\Phi_m'(y')$ on the exit lead for outgoing modes
\begin{equation}  
t_{nm}=2i(k_nk_m')^{1/2}\int dydy'\Phi_n(y)G^+(x,y|x',y')\Phi_m'(y'), 
\label{trans} 
\end{equation}  
and $x$ and $x'$ lie anywhere on the entrance side and  exit side, 
respectively (See Fig. 1a-c).
(From here on we will use units such $\hbar=2m=1$.) Reflection between two 
modes $n$ and 
$m$ on the same side is given by 
\begin{equation}  
r_{nm}=\delta_{nm}-2i(k_nk_m)^{1/2}\int dydy'\Phi_n(y)G^+(x,y|x,y')\Phi_m(y'), 
\label{refl} 
\end{equation}  
where $x$ lies anywhere on the entrance side.  For open channels $n$ of the 
entrance side the transmission and reflection amplitudes fulfill the sum 
rules 
\begin{equation}  
\sum_m |t_{nm}|^2+\sum_{m'}|r_{nm'}|^2=1, 
\end{equation}  
where the two summation goes for channels $m$ on the exit side and $m'$ 
on the entrance side. These are the consequences of the 
probability and current conservation.  Thanks to these relations, in 
infinitely long wave guides, the conductance can be expressed with the 
reflection coefficients 
\begin{equation}  
G(E_F)=\frac{2e^2}{h}\left(N-\sum_{m,n}|r_{nm}|^2\right), 
\label{condrefl} 
\end{equation}  
where $N$ is the number of open channels on the exit side.  
  
The direct connection between the Green function and the conductance  
makes it possible to develop a semiclassical approximation for the  
conductance based on the semiclassical Van Vleck - Gutzwiller type  
expressions for the Green function. Next, we summarize the main tools  
of the semiclassical description.  
  
\section{The semiclassical Green function}  
 
The Schr\"odinger equation in the waveguides leads to the  
Helmholtz equation  
\begin{equation}  
(\Delta+k^2)\psi(q)=0,  
\end{equation}   
where the notation $k=\sqrt{2mE/\hbar^2}$ has been introduced.  
The wavefunction vanishes on the boundary as we discussed above.  
The energy domain Green function of the Helmholtz equation is  
defined by  
\begin{equation}  
(\Delta +k^2)G(q,q',E)=\delta(q-q'),  
\end{equation}  
and $G(q,q',E)$ should vanish on the boundary of the waveguide too.  
In absence of walls in two dimensions the outgoing Green function   
is given by   
\begin{equation}  
G^+(q,q',E)=-\frac{i}{4}H^{(1)}_0(kd(q,q')),  
\end{equation}  
where $H^{(1)}_0(x)$ is the Hankel function of first kind and  
$d(q,q')$ is the distance between $q$ and $q'$.  
The semiclassical ($\hbar\rightarrow 0$) approximation of this  
free space Green function can be recovered from the asymptotic 
form of the Hankel function for large argument  
\begin{equation}  
G^+(q,q',E)=\frac{1}{\sqrt{8\pi kd(q,q')}}e^{ikd(q,q')-i3\pi/4}.  
\end{equation}   
A very useful optical interpretation of this formula can be
given  by tracing the ray connecting  $q$ with $q'$.  
The phase of the Green function is the classical action  
calculated along the ray $$\int_{q}^{q'} p(q'')dq''=kd(q,q').$$ 
The amplitude of the  
Green function is the square root of the intensity  
\begin{equation}  
I(r)= \frac{1}{8\pi kr} \label{intens}  
\end{equation}  
of a radiating point source where the distance from the source is $r=d(q,q')$.  
The phase factor $e^{-i3\pi/4}$ is the Maslov index of the caustics  
singularity standing right in the source point $q$.  

In the presence of walls, in general, more than one ray from $q$ can reach  
the point in $q'$. Then the semiclassical Green function is a  
sum for the raywise Green functions calculated along the rays  
\begin{equation}  
G(q,q',E)=\sum_{\forall r \\ q\rightarrow q'} G_{0}^r(q,q',E).  
\end{equation}  
The phase of the Green functions $G_{0}^r$ is given by the classical action   
plus the Maslov index including the caustics in the source point.  
The amplitude is the square root of the intensity observable in $q'$  
coming form $q$ along the ray. In the  
endpoint we will see the mirror image of the source trough  
perfect mirrors given by the walls. The intensity in this  
case is given by the same formula (\ref{intens}), but  
now the effective radius of the spherical wave  
coming from $q$ should be used. If the walls are non curved  
this radius is simply the distance $d_r(q,q')$ of $q$ and $q'$ along the ray.  
In this case the Green function is given by  
\begin{equation}  
G_0^r(q,q',E)=  
\frac{(-1)^{n_r}}{\sqrt{8\pi kd_r(q,q')}}e^{ikd_r(q,q')-i3\pi/4},  
\label{gnoncurved} 
\end{equation}  
where $n_r$ is the number of bounces along the ray.  
For the semiclassical expression of the Green function in presence of 
curved walls the effective radius is given \cite{Vattay3} by the product 
of the distance between the source and the first bounce point $l_1$ times 
the stretching factor $\Lambda$ 
\begin{equation} 
r_{eff}=l_1\Lambda. 
\end{equation}  
The stretching factor is the analog of the optical magnification factor of 
curved mirrors. It can be calculated as 
\begin{equation} 
\Lambda=\prod_{i=1}^{n-1} (1+\kappa_i^+ l_i), 
\end{equation} 
where $l_i$ is the length of the free flight between the $i^{th}$ and 
the $(i+1)^{th}$ bounces and $\kappa$ is the Bunimovich-Sinai 
curvature. The Bunimovich-Sinai curvature is defined recursively 
\begin{eqnarray} 
\kappa_{i}^+&=&\kappa_i^-+2/r_i\cos(\phi_{i}), \nonumber \\ 
\kappa_{i+1}^-&=&\kappa_{i}^+/(1+\kappa_i^+ l_i), 
\end{eqnarray} 
where $r_i$ is the radius of curvature of the wall in the 
point of bounce and $\phi_i$ is the angle of incidence. 
The initial condition of the recursion is $\kappa_1^-=1/l_1$, 
which is the Bunimovich-Sinai curvature of the wavefront 
arriving at the first bounce point from the source.  
The Green function reads as 
\begin{equation}  
G_0^r(q,q',E)=  
\frac{(-1)^{n_r}}{\sqrt{8\pi k l_1 \Lambda_r}}e^{ikd_r(q,q')-i3\pi/4}. 
\label{gcurved} 
\end{equation}  
In the special case of non-curved walls the stretching factor 
(optical magnification) is $\Lambda=d_r(q,q')/l_1$ yielding  
the expression (\ref{gnoncurved}). 
 
\section{Geometric Theory of Diffraction}  
 
The semiclassical approximation works well when changes of the 
potential are on a length scale much larger than the wavelength 
of the electron $1/k$ \cite{csoport3}. 
Even if the manufactured system in large 
satisfies this condition, there are isolated points such as corners or  
impurities where this cannot be the case. These points require 
special treatment \cite{Vattay1,Pavloff,Uzy}. They break the smooth wavefronts of 
the semiclassical wave propagation and create diffracted waves  
coiling out from them. Such diffractive 
points can be considered as new wave sources whose strength 
is proportional with the strength of the incident wave. 
Their contribution of such a specific ray to the Green function is 
given by 
\begin{equation} 
G_d(q,q',k)=G_0(q,q_0,k){\cal D}(\phi_{in},\phi_{out},k)G_0(q_0,q',k), 
\label{difgreen} 
\end{equation} 
where $G_0(q,q_0,k)$ is the  Green function calculated along the 
ray connecting the source point $x$ and the diffractive point 
$q_0$, ${\cal D}(\phi^r_{in},\phi^r_{out})$ is the diffraction constant 
which can depend on the incoming and the outgoing 
angle of the ray and the energy. $G_0(q_0,q',k)$ is the Green function 
calculated from the source to the point of observation $q'$. 
If there is more than one ray connecting $q$ with $q_0$ and $q_0$ 
with $q'$ then each ray configuration from $q$ to $q'$ will 
contribute to the Green function according to (\ref{difgreen}). 
 
The diffraction constant can be determined for different  
geometric situations. For wedge diffraction 
\begin{equation}
{\cal D}(\phi_{in},\phi_{out})=\frac{\sin(\pi/n)}{n}\left[
\frac{1}{\cos(\pi/n)-\cos((\phi_{in}-\phi_{out})/n)}-
\frac{1}{\cos(\pi/n)-\cos((\pi+\phi_{in}+\phi_{out})/n)}\right],
\end{equation}
where $(2-n)\pi$ is the opening angle of the wedge. For details we refer to
Ref. \cite{Keller1,Vattay3,Pavloff}. For diffraction on impurities the $s$ wave scattering is
dominant and the diffraction constant ${\cal D}$ is isotropic, depending
only on the energy. Typical example is a $\delta$-shaped potential with 
scattering strength ${\cal D}$.

\section{Semiclassical transmission and reflection} 
 
Semiclassically the 
transmission (\ref{trans}) and reflection (\ref{refl}) matrix elements 
are calculated
by replacing the Green function in (\ref{trans}) and (\ref{refl}) with 
the semiclassical expressions (\ref{gcurved}) 
or (\ref{gnoncurved}). The wave functions $\Phi_n(y)$ and 
$\Phi_m'(y')$ can be decomposed to exponentials 
\begin{eqnarray} 
\Phi_n(y)&=&\frac{1}{2i}\sqrt{\frac{2}{W}} 
\left(e^{in\pi y/W}-e^{-in\pi y/W}\right), \nonumber \\
\Phi_m(y')&=&\frac{1}{2i}\sqrt{\frac{2}{W'}} 
\left(e^{i m\pi y'/W'}-e^{-i m\pi y'/W'}\right)\label{sinexp}. 
\end{eqnarray} 
The transmission and reflection matrix elements are 
then sums of four subintegrals 
\begin{eqnarray} 
r_{nm}&=&r_{nm}^{++}+r_{nm}^{+-}+r_{nm}^{-+}+r_{nm}^{--}, \nonumber \\ 
t_{nm}&=&t_{nm}^{++}+t_{nm}^{+-}+t_{nm}^{-+}+t_{nm}^{--},
\end{eqnarray} 
where 
\begin{eqnarray} 
t_{nm}^{\pm\pm}&=&i(k_nk_m')^{1/2}\frac{1}{\sqrt{WW'}}\int dydy'
e^{\pm i n\pi y/W\pm i 
m \pi y'/W'-i\nu\pi/2}G^+(x,y|x',y'), \nonumber \\ 
r_{nm}^{\pm\pm}&=&\frac{\delta_{nm}}{4}-i(k_n k_m)^{1/2}\frac{1}{W}
\int dydy'e^{\pm i n\pi y/W\pm i 
m \pi y'/W-i\nu\pi/2}G^+(x,y|x,y'),\label{trpm}
\end{eqnarray}  
where $\nu=\pm1\pm1$. 
By using (\ref{gcurved}) 
or (\ref{gnoncurved}) in semiclassical approximation one can calculate these 
integrals with saddle point approximation. The saddle point conditions 
for various 
$t_{nm}^{\pm\pm}$-s are 
\begin{eqnarray} 
k\partial_y d_r(x,y|x',y')&=&\mp \pi n / W \qquad \mbox{and} \nonumber \\ 
k\partial_{y'} d_r(x,y|x',y')&=&\mp \pi m / W', 
\end{eqnarray} 
and for  $r_{nm}^{\pm\pm}$-s are 
\begin{eqnarray} 
k\partial_y d_r(x,y|x,y')&=&\mp \pi n / W \qquad\mbox{and} \nonumber \\ 
k\partial_{y'} d_r(x,y|x,y')&=&\mp \pi m / W. 
\end{eqnarray} 
If we introduce the initial $\phi_{in}$ and final $\phi_{fin}$  
angle of the rays with respect to 
the direction of the leads these conditions are equivalent
with selecting rays whose initial angle is determined by 
\begin{equation} 
\sin(\phi_{in})=\mp\frac{n\pi}{kW} 
\end{equation} 
and their final angle is 
\begin{eqnarray} 
\sin(\phi_{fin})&=&\mp\frac{m\pi}{kW'} \qquad\mbox{for $t_{nm}$}, \nonumber \\ 
\sin(\phi_{fin})&=&\mp\frac{m\pi}{kW} \qquad\mbox{for $r_{nm}$}, 
\end{eqnarray} 
respectively.  
 
For irregularly shaped wave guides one can in general 
find rays starting and ending with given initial 
and final angles. We can call such systems "chaotic". 
In wave guides formed by regularly shaped walls it is possible 
that in the bouncing process a  
components of the momentum is conserved in addition to the energy conservation.
In these systems only  one of the saddle point conditions can be satisfied. 
We can call them  "integrable".  One example is the waveguide on Fig. 1a. 
There are systems, where chaotic and integrable behaviour 
coexist. For certain initial angle one can find rays to any 
final angle, while for other initial conditions not. 
These systems are "mixed".  The conductance of chaotic, integrable  
and mixed systems are relatively well understood.
  
The simple, typically human designed systems of Fig. 1b and Fig. 1c  
do not belong to any of these categories. We may call them "pseudo integrable" 
systems. These systems are not chaotic, since there exist is no ray in 
general connecting a given initial and a final angle. An initial angle 
allows only a finite number of possible final angles. So, they are more 
like integrable systems in this respect. However, when we take into 
account the diffractive points, rays can leave them in any direction. 
So, we can always find diffractive rays connecting any initial 
angle with any final one. In this respect, these systems are more 
chaotic like. 
 
In the wave guides on Fig. 1b and 1c starting from the left with a 
given initial angle and bouncing on the walls, the electron can end up on  
the right with a definite final angle. Return to the left is impossible. 
Except for a classically "measure zero" fraction of rays  hitting 
the impurity or the wedge. The contribution of these rays to the transmission 
and to the reflection 
can be computed by substituting the diffractive part of the Green
function (\ref{difgreen}) into (\ref{trpm}). 
The length of the ray connecting $(x,y)$ with $(x',y')$ or 
$(x,y')$ via the diffractive point $(x_0,y_0)$ is 
\begin{eqnarray} 
d(x,y|x',y')&=&d(x,y|x_0,y_0)+d(x_0,y_0|x',y'), \nonumber \\ 
d(x,y|x,y')&=&d(x,y|x_0,y_0)+d(x_0,y_0|x,y'). 
\end{eqnarray} 
Since the position of the diffractive point $(x_0,y_0)$ is fixed, the  
saddle point conditions for the initial and final segments decouple 
for $t_{nm}^{\pm\pm}$-s 
\begin{eqnarray} 
k\partial_y d(x,y|x_0,y_0)&=&\mp \pi n / W \qquad \mbox{and} \nonumber \\ 
k\partial_{y'} d(x_0,y_0|x',y')&=&\mp \pi m / W', 
\label{sad1}
\end{eqnarray} 
and for the $r_{nm}^{\pm\pm}$-s 
\begin{eqnarray} 
k\partial_y d(x,y|x_0,y_0)&=&\mp \pi n / W \qquad\mbox{and} \nonumber \\ 
k\partial_{y'} d(x_0,y_0|x,y')&=&\mp \pi m / W. \label{sad2}
\end{eqnarray} 
The saddle point integrals 
\begin{eqnarray} 
t_{nm}^{\pm\pm}&=&\frac{i(k_nk_m')^{1/2}}{\sqrt{WW'}}\int dydy'
e^{\pm i n\pi y/W\pm i 
m \pi y'/W'-i\nu\pi/2}G_0(x,y|x_0,y_0)
{\cal D}(\phi_{in},\phi_{out})G_0(x_0,y_0|x',y'), \nonumber \\ 
r_{nm}^{\pm\pm}&=&-\frac{i(k_n k_m)^{1/2}}{W}
\int dydy'e^{\pm i n\pi y/W\pm i 
m \pi y'/W-i\nu\pi/2}G_0(x,y|x_0,y_0)
{\cal D}(\phi_{in},\phi_{out})G_0(x_0,y_0|x,y')\label{dtrpm} 
\end{eqnarray}
then can be carried out separately. The Kronecker delta disappears, as 
usual \cite{BarSto,csoport3}, 
due to the trivial direct rays from $y$ to $y'$.
Assuming straight walls we get
\begin{eqnarray} 
t_{nm}^{\pm\pm}&=&-i{\cal D}(\phi_{in},\phi_{out}) 
\frac{(-1)^{n_r}}{4\sqrt{WW'k_nk_m'}}e^{ik(d(x,\bar{y}|x_0,y_0)+d(x_0,y_0|x',
\bar{y}')) 
\pm i n\pi \bar{y}/W\pm i 
m \pi \bar{y}'/W'-i\nu\pi/2},  \nonumber \\ 
r_{nm}^{\pm\pm}&=&+i{\cal 
D}(\phi_{in},\phi_{out})\frac{(-1)^{n_r}}{4W\sqrt{k_nk_m}} 
e^{ik(d(x,\bar{y}|x_0,y_0)+d(x_0,y_0|x,\bar{y}'))\pm i n\pi \bar{y}/W\pm 
im \pi 
\bar{y}'/W-i\nu\pi/2}, 
\end{eqnarray}  
where $n_r$ is the total number of bounces on the straight wall
and $\bar{y}$ and $\bar{y}'$ are the solutions for the saddle
point conditions (\ref{sad1}) and (\ref{sad2}). The $\pm$ signs in
$\bar{y}_{\pm}$ and $\bar{y}_{\pm}'$ have been suppressed for brevity.
 
A ray with only one diffractive point is the most elementary 
possibility for diffraction. More complicated diffractive rays are 
those, which return to the diffractive point several times.  These 
events are described by the following Green function 
\begin{equation} 
G_d(x,y|x',y')=G_0(x,y|x_0,y_0)G_0(x_0,y_0|x',y') 
\left(  
{\cal D}(\phi_{in},\phi_{out}) + \sum_{r} 
{\cal D}(\phi_{in},\phi_r) 
G_0^r(\phi_r,\phi_r'){\cal D}(\phi_r',\phi_{out})\right), 
\end{equation}  
where $\phi_{in}$ and $\phi_{out}$ denote the angles under 
which the incoming and outgoing rays reach first and leave finally 
the diffractive point, $\phi_r$ and $\phi_r'$ stand for  
starting and ending angles of ray loops starting and ending in 
the diffractive points and $G_0^r(\phi_r,\phi_r')$ denotes the Green function 
computed along this ray loop. The expression in the bracket 
can be considered as an effective diffraction constant, renormalized  
by the interaction with the environment. The closed ray loops starting 
and ending in a diffractive point were introduced in Ref.\cite{Vattay1} 
first and are called diffractive periodic orbits. The summation for 
$r$ is a summation for all the possible diffractive periodic orbits 
starting and ending in the diffractive point 
\begin{equation} 
\tilde{{\cal D}}(\phi_{in},\phi_{out})=  
{\cal D}(\phi_{in},\phi_{out}) + \sum_{p} A_pe^{ikd_p}, 
\label{reno} 
\end{equation} 
where $d_p$ is the length of the 
diffractive periodic orbit and $A_p=|{\cal D}(\phi_{in},\phi_p,k) 
G_r(\phi_p,\phi_p'){\cal D}(\phi_p',\phi_{out})|$.   
The transmission and reflection matrix elements then can be
written in terms of the renormalized diffraction constants 
\begin{eqnarray} 
t_{nm}^{\pm\pm}&=&-i\tilde{{\cal D}}(\phi_{in},\phi_{out}) 
\frac{(-1)^{n_r}}{4\sqrt{WW'k_nk_m'}}e^{ik(d(x,\bar{y}|x_0,y_0)+
d(x_0,y_0|x',\bar{y}')) 
\pm i n\pi \bar{y}/W\pm i 
m \pi \bar{y}'/W'-i\nu\pi/2},  \nonumber \\ 
r_{nm}^{\pm\pm}&=&+i\tilde{{\cal 
D}}(\phi_{in},\phi_{out})\frac{(-1)^{n_r}}{4W\sqrt{k_nk_m}} 
e^{ik(d(x,\bar{y}|x_0,y_0)+d(x_0,y_0|x,\bar{y}'))\pm i n\pi \bar{y}/W\pm im 
\pi \bar{y}'/W-i\nu\pi/2}.\label{dptrpm} 
\end{eqnarray} 
The expression derived here is the main result of the  paper and 
expresses the diffractive part of the transmission and reflection 
matrix elements in terms of diffractive periodic orbits and the 
rays reaching and leaving the impurity. 
 
In case of more than one diffractive points $q_0,...,q_k$,  
one can imagine more complicated situations. The ray from 
$y$ can reach $q_l$ first, then jumps  to $q_s$ and goes 
to $y'$. Beyond the direct ray between $l$ and $s$, there are rays which 
start in $l$ and reach $s$ in different ways, possibly trough  
diffraction on other diffractive points or trough bouncing  
on the walls or trough some combination of these.  
These type of processes are described by the 
Green function 
\begin{equation} 
G_d(x,y|x',y')=\sum_{l,s}G_0(x,y|x_l,y_l)G_0(x_s,y_s|x',y') 
\left( \sum_{r} {\cal D}(\phi_{y\rightarrow l},\phi_r) 
G_0^r(\phi_r, \phi_r'){\cal D}(\phi_r',\phi_{s\rightarrow y'}) \right), 
\end{equation}  
where $G_r(\phi_r, \phi_r')$ is the Green function computed along 
the ray leaving $l$ at angle $\phi$ and reaching $s$ at angle $\phi'$. 
We can introduce the notations 
\begin{equation} 
\tilde{{\cal D}}_{ls}(\phi_{y\rightarrow l},\phi_{s\rightarrow y'},k)= 
 \sum_{r} 
{\cal D}(\phi_{y\rightarrow l},\phi_r) 
G_0^r(\phi_r, \phi_r'){\cal D}(\phi_r',\phi_{s\rightarrow y'}), 
\end{equation}  
where the diagonal terms  
$\tilde{{\cal D}}_{ll}(\phi_{y\rightarrow l},\phi_{s\rightarrow y'},k)$ 
are defined 
as the renormalized diffraction constants defined in (\ref{reno}). 
Finally, we can compute the transmission and reflection matrix elements 
as before and get 
\begin{eqnarray} 
t_{nm}^{\pm\pm}&=&-i\sum_{ls}\tilde{{\cal D}}_{ls}(\phi_{in,l},\phi_{out,s},k) 
\frac{(-1)^{n_{ls}}}{4\sqrt{WW'k_nk_m'}}e^{ik(d(x,\bar{y}|x_l,y_l)+
d(x_s,y_s|x',\bar{y}')) 
\pm i n\pi \bar{y}/W\pm i  
m \pi \bar{y}'/W'-i\nu\pi/2}, \nonumber \\ 
r_{nm}^{\pm\pm}&=&+i\sum_{ls}\tilde{{\cal D}}_{ls}(\phi_{in,l},\phi_{out,s}) 
\frac{(-1)^{n_{ls}}}{4W\sqrt{k_nk_m}}e^{ik(d(x,\bar{y}|x_l,y_l)+
d(x_s,y_s|x',\bar{y}')) 
\pm i n\pi \bar{y}/W\pm i 
m \pi \bar{y}'/W-i\nu\pi/2}.\label{dmtrpm} 
\end{eqnarray} 
This is the most general expression, within the framework of the GDT, 
for the transmission and reflection matrix elements.  
 
Next, we are going to apply the theory developed here for the 
examples announced in the introduction and show which are 
the consequences of diffractive contributions for the conductance 
of these systems.  
 
\section{Erosion of conductance steps} 
 
We have calculated the conductance of an infinitely long 
waveguide with a point like impurity in its center and the 
broken-strip waveguide (see Fig. 2). 
\begin{figure}
\centerline{\strut\psfig{figure=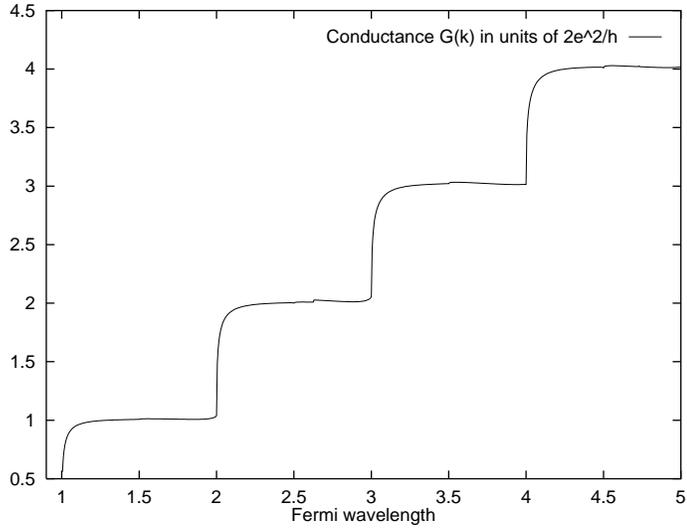,width=70mm}}
\centerline{\strut\psfig{figure=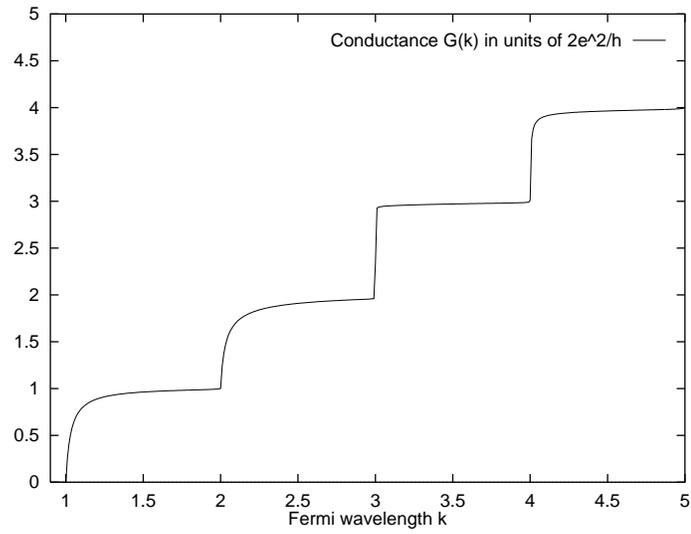,width=70mm}}
\caption{\small Conductance as a function of the Fermi wavelength in units of 
$\frac{\pi}{W}$
{\bf a,} For the infinite strip with impurity (${\cal D}=1$). {\bf b,} 
For the broken-strip waveguide. The angle of the
strips measured from the
line connecting the two wedges were $\pi/6$ and $\pi/2$.}
\end{figure}
The details 
of the calculation will be published elsewhere\cite{CsPSzV,CsV}. 
The geometry in both cases is such, that a classical electron starting  
from the left of the device cannot return back via classical 
bounces on the walls\cite{zero}.  
The usual semiclassical theory, in the absence of returning rays, predicts no 
reflection for these systems and the semiclassical reflection 
matrix elements $r_{nm}$ all vanish. According to (\ref{condrefl}) 
the conductance of such systems should coincide 
with  that of an ideal straight waveguide $G(E_F)=\frac{2e^2}{h} N$. As we 
can see on Fig. 2, this is not the case. 
The  
difference between the sharp ideal staircase and the actual 
conductance is the consequence of the non-vanishing reflection. 
In case of the pointlike impurity the reflection can be determined 
exactly and various terms can be interpreted from a diffractive 
point of view. The exact reflection matrix elements for an 
impurity in an infinite straight waveguide are 
\begin{equation} 
r_{nm}=-\Phi_n(y_0)\Phi_m(y_0)\frac{1}{2i\sqrt{k_nk_m}}e^{ik_n|x-x_0|+ 
ik_m|x-x_0|}\frac{{\cal D}}{1-{\cal D}\tilde{G}_E(x_0,y_0|x_0,y_0)}, 
\end{equation} 
where the exact Green function of the empty guide 
\begin{equation} 
G_E(x,y|x',y')=\sum_n\frac{\Phi_n(y)\Phi_n(y')}{2ik_n}e^{ik_n|x-x'|} 
\end{equation} 
has been evaluated on the impurity and the singularity has been  
removed 
\begin{equation} 
\tilde{G}_E(x_0,y_0|x_0,y_0)=\lim_{\epsilon\rightarrow 0} 
\left(G_E(x_0+\epsilon,y_0|x_0,y_0)+\frac{1}{4}Y_0(k\epsilon)\right),
\end{equation}
where $Y_0(x)$ is the Neumann function.  
By using (\ref{sinexp}) we can determine the exact $r^{\pm\pm}_{nm}$ 
elements: 
\begin{equation} 
r_{nm}^{\pm\pm}=e^{\pm iny_0/W \pm imy_0/W}\frac{i}{4W\sqrt{k_nk_m}} 
e^{ik_n|x-x_0|+ 
ik_m|x-x_0|-i\nu\pi/2}\frac{{\cal D}}{1-{\cal D}\tilde{G}_E(x_0,y_0|x_0,y_0)}. 
\label{dirr} 
\end{equation} 
These terms can be interpreted as contributions from various rays 
reaching the impurity. For example a ray starting in $x,\bar{y}$ under angle 
$\sin(\phi_{in})=n\pi/Wk$ will reach $x_0,y_0$ by flying 
a distance $|x-x_0|$ in horizontal direction and $|y_0-\bar{y}|+ rW$ 
in the vertical direction, where $r$ is the number of bounces 
during the flight, the total length of the ray is  
$$d(x,\bar{y}|x_0,y_0)=(|x-x_0|^2 +(|y_0-\bar{y}|+ rW)^2)^{1/2},$$ 
and 
$$\cos(\phi_{in})=|x-x_0|/d(x,\bar{y}|x_0,y_0),\;\;\sin(\phi_{in})=
(|y_0-\bar{y}|+rW)/d(x,\bar{y}|x_0,y_0).$$ 
Using the relation $k_n=k\cos(\phi_{in})$ the full exponential 
can be written as 
$$k_n|x-x_0|+n\pi y_0/W=kd\cos^2(\phi_{in})+
k\sin(\phi_{in})(d\sin(\phi_{in})- rW+\bar{y})=
kd-rn\pi+n\pi\bar{y}/W.$$
Replacing the phase terms in (\ref{dirr}) with this and analogous 
expression for the other ray (see Fig.\ 3a) we recover (\ref{dptrpm}) exactly with renormalized 
diffraction constant 
\begin{equation} 
\tilde{{\cal D}}=\frac{{\cal D}}{1-{\cal D}\tilde{G}_E(x_0,y_0|x_0,y_0)}. 
\end{equation} 
It is very reassuring to see diffraction theory reproducing the exact result 
in this case. 
\begin{figure}
\centerline{\strut\psfig{figure=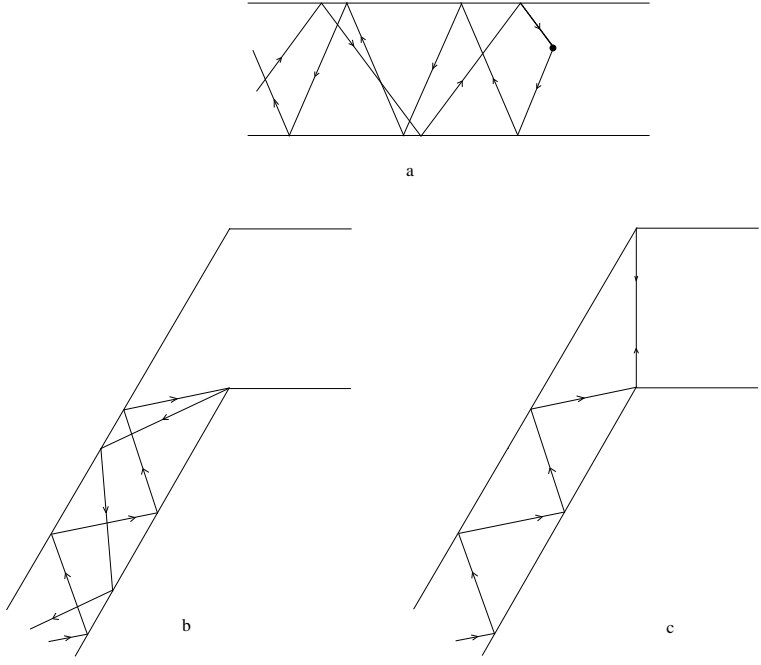,width=100mm}}
\caption{\small Typical diffractive rays. {\bf a,} Primary rays reaching
and leaving the diffractive point created by the pointlike
impurity. {\bf b} Primary rays reaching
and leaving the diffractive point in the wedge. {\bf c}
Primary ray and the ray bouncing forth and back between
the diffractive points.} 
\end{figure}

The regularized Green function can be written semiclassically as a 
sum for all trajectories starting on and returning to the singularity\cite{reg}
\begin{equation}  
\tilde{G}_E(q_0,q_0)\approx\sum_{r\;q_0\rightarrow q_0}  
\frac{(-1)^{n_r}}{\sqrt{8\pi k d_r}}e^{ikd_r-i3\pi/4}.
\end{equation}  
Accordingly the renormalized diffraction constant  
\begin{equation} 
\tilde{{\cal D}}={\cal D}+{\cal D}^2\tilde{G}_0(x_0,y_0|x_0,y_0)+ 
{\cal D}^3\tilde{G}_0^2(x_0,y_0|x_0,y_0)+ ... 
\end{equation} 
can be viewed as a sum for all returning rays having one, two, ... etc. 
diffractions on their way. 
 
It is very enlighting to study on this simple model how diffraction 
erodes the conductance steps close to Fermi wavelengths where 
new channels are opened $k\rightarrow \pi n/W$ ($k_n\rightarrow 0)$.  
Around these
Fermi wavelengths the regularized Green function is dominated by a single 
term 
\begin{equation} 
\tilde{G}_0(q_0,q_0)\approx\frac{\Phi_n(y_0)\Phi_n(y_0)}{2ik_n}. 
\end{equation} 
The squares of the reflection matrix elements can be approximated 
by 
\begin{equation} 
|r_{nm}|^2=\frac{k_n}{k_m}\frac{|\Phi_m(y_0)|^2}{|\Phi_n(y_0)|^2}
\frac{1}{1+\frac{4k_n^2}{{\cal D}^2|\Phi_n(y_0)|^4}}.
\end{equation} 
The nondiagonal 
reflection coefficients are small and vanish for $k_n\rightarrow +0$, 
while the diagonal is a Lorentzian of width $\Delta_n={\cal D}|\Phi_n(y_0)|^2/2$ 
\begin{equation} 
|r_{nn}|^2=\frac{1}{1+(k_n/\Delta_n)^2},\qquad k_n>0.
\end{equation} 
Accordingly, the conductance steps have an eroded shape 
\begin{equation} 
G=\frac{2e^2}{h}(N-|r_{nn}|^2)=\frac{2e^2}{h}\left(N-1+\frac{(k_n/\Delta_n)^2} 
{1+(k_n/\Delta_n)^2}\right). 
\end{equation} 
We can see, that the average width of the erosion of the conductance 
steps is a primarily information about the level of diffraction
in a waveguide (see Fig.\ 2a).

The diffraction theory is not exact for the broken-strip waveguide, but the 
conductance shows qualitatively the 
same behaviour (Fig.\ 2b). The main difference is that the diffraction  
constants here depends on the angle and reflection 
matrix elements are more complicated. The primary diffractive rays reaching 
and  leaving the wedge are depicted on Fig.\ 3b. For $k$ just above 
$\pi n/W$ the initial angle is close 
to $\phi_{in}\approx\pi/2$ e.g. the electron is bouncing almost 
perpendicularly 
on the walls of the guide and cannot reach the upper 
wedge via bounces. On Fig.\ 3c a ray reaching the lower wedge by bounces 
and then going to the other wedge by diffraction is shown. When $k$ gets 
a bit larger, the initial angle decreases and rays reaching the 
other wedge directly should be considered. The sudden appearance of 
the new ray causes an abrupt change in $r_{11}$  at $k=1.5$ on Fig.\ 4.
 
For larger $k>>\pi n/W$ Fermi wavelengths the reflection matrix elements 
decrease rapidly and show oscillations small amplitude oscillations 
dominated by the oscillations coming from individual diffractive 
orbits (see Fig.\ 4).
\begin{figure}
\centerline{\strut\psfig{figure=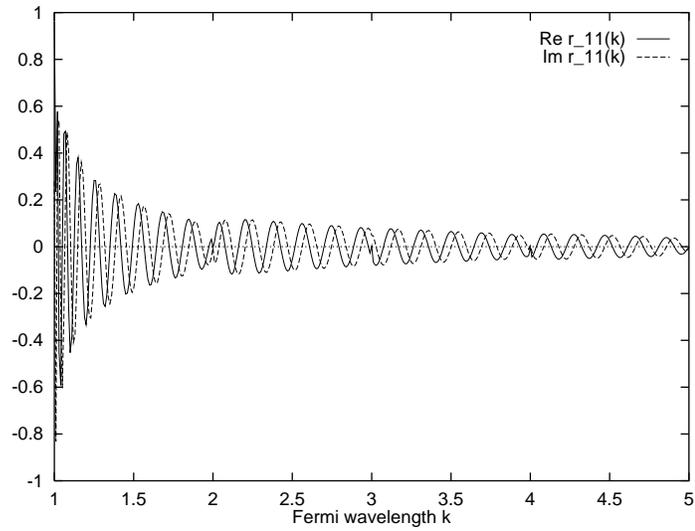,width=70mm}}
\centerline{\strut\psfig{figure=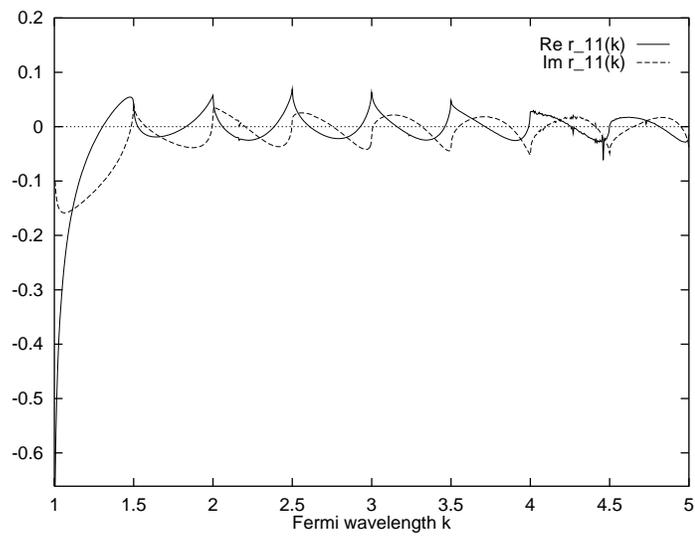,width=70mm}}
\caption{\small The reflection matrix element $r_{11}$ as a 
function of the Fermi wavelength in units of 
$\frac{\pi}{W}$ {\bf a,} for the
strip waveguide with impurity {\bf b,} for the
broken-strip waveguide. (Re $r_{11}(k)$, Im $r_{11}(k)$ 
are denoted by solid and dashed lines, respectively.)}
\end{figure}

\section{Diffractive periods in the power spectrum} 
 
For $k>>\pi n/W$ both $k_n$ and $k_m$ can be approximated with $k$. 
The primary diffractive rays are almost parallel with the walls of 
the guide $\phi\approx 0$. Since the Green functions of the rays have a 
prefactor 
proportional with $1/\sqrt{k}$, multiple diffractions are largely 
suppressed. The reflection matrix elements can be well approximated  
in leading $1/k$ order by keeping rays with one and two diffractive
points only.  Rays going trough two impurities while
traversing the system are the most trivial examples. The first non-trivial 
contributions come from rays reaching a diffractive point and leaving
it after making a loop.
 
In case of the point impurity this means that we keep terms proportional 
with the first and second power of the diffraction constant only: 
\begin{equation} 
r_{nm}\approx-\Phi_n(y_0)\Phi_m(y_0)\frac{1}{2ik}e^{ik_n|x-x_0|+ 
ik_m|x-x_0|}\left({\cal D}+{\cal D}^2\tilde{G}_E(x_0,y_0|x_0,y_0)\right). 
\end{equation} 
The regularized Green function can be written as a sum for rays starting 
and ending on the impurity yielding 
\begin{equation} 
r_{nm}\approx-\Phi_n(y_0)\Phi_m(y_0)\frac{1}{2ik}e^{i2k|x-x_0|} 
\left({\cal D}+{\cal D}^2\sum_{q_0\rightarrow q_0}  
\frac{(-1)^{n_r}}{\sqrt{8\pi k d_r}}e^{ikd_r(q_0,q_0)-i3\pi/4}\right). 
\end{equation} 
Beyond the trivial phase factor $e^{i2k|x-x_0|}$ coming from the 
ray reaching and leaving the impurity, the most important 
oscillations are caused by rays making a single loop starting 
and ending on the impurity. The length of these orbits can be 
calculated and can be grouped in three subcategories   
\begin{eqnarray} 
d_r&=&2y_0+2Wr, \nonumber \\ 
d_r&=&2(W-y_0)+2Wr,\\ 
d_r&=&2Wr, \nonumber
\end{eqnarray} 
where $r=0,1,2,...$. The amplitude of the 
oscillations is proportional with $1/\sqrt{d_r}\sim 1/\sqrt{r}$. 
  
In the broken-strip waveguide the situation is somewhat more complicated. 
Loops starting and ending on the wedges can do it typically by making 
additional diffraction on the other wedge as shown on Fig. 2. 
But this diffraction process is somewhere in between a usual 
perpendicular bounce back from a wall and a diffraction in the 
corner. Therefore, the contributions of these orbits is not  
suppressed after one bounce. To see how the amplitude of these 
orbits decreases, we Fourier transformed the large $k$ tail 
of the reflection matrix element $r_{11}$ and calculated 
the power spectra 
\begin{equation} 
\tilde{r}_{11}(l)=\left|\int_{k_0}^{\infty} dk r_{11}(k)e^{-ikl}\right|, 
\end{equation} 
where $k_0\approx 2\pi/W$ has been chosen.
We have multiplied $r_{11}$ with a phase factor $e^{i2kL}$ 
in order to remove the main oscillation coming from the distance 
$L$ between the lower wedge and the starting point of our coordinate 
system.
On Fig. 5 we can see that the contribution of the suspected orbits 
is the most significant. 
\begin{figure}
\centerline{\strut\psfig{figure=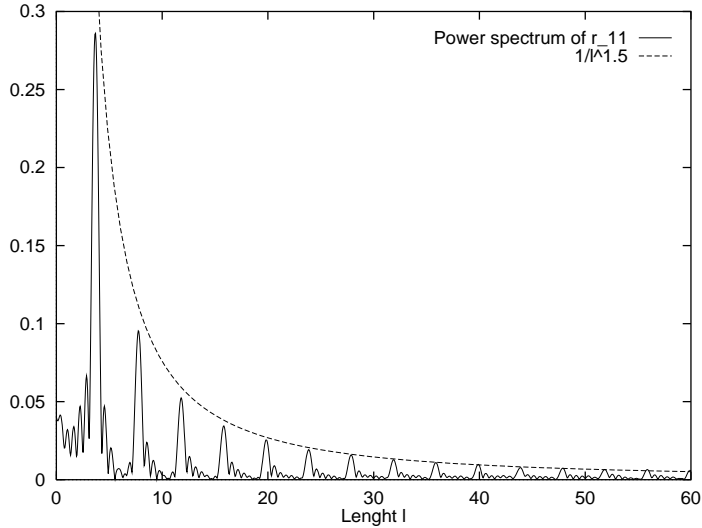,width=70mm}}
\caption{\small The power spectrum of the reflection matrix element
$r_{11}$ for the broken-strip waveguide. The heights of the peaks 
are proportional with the $l^{-1.5}$.}
\end{figure}
Their amplitude decay as 
$\sim 1/r^{3/2}$, which is faster than it is in the impurity 
case but slower than the exponential decay predicted for multiple 
diffraction. 
 
We can conclude, that in these two pseudo integrable situations 
the non-trivial part of the power spectra of the reflection 
matrix elements shows a $\sim 1/l^\beta$ type 
scaling with $\beta=1/2$ and $\beta=3/2$. In a similar but 
chaotic situation the power spectrum decays much faster 
due to the exponential suppression of the amplitude of long orbits 
in the semiclassical Green function. The special half integer decays 
can probably be always associated with the presence of impurity 
and wedge diffraction.

\section{Summary} 
In this paper we analyzed the conductance of mesoscopic devices 
where diffraction plays a major role. Such pseudo integrable 
systems are typical, when the form of the device is human 
designed. We derived formulas expressing the reflection and transition 
matrix elements for one and many diffractive points. We verified 
this for the special case of impurity and wedge diffraction.  
Diffraction can cause back scattering in situations, where usual 
semiclassical back scattering is absent causing an  erodation 
of ideal conductance steps. The length of diffractive periodic orbits  
and diffractive loops can be detected in the power spectrum of the 
reflection matrix elements. The large $l$ part of the power spectrum 
shows $1/l^{1/2}$ decay due to impurity scattering and $1/l^{3/2}$ 
decay due to wedge scattering which we think is a universal sign 
of the presence of diffractive scattering in pseudo integrable  
waveguides.

\section{Acknowledgments}
We would like to thank K. Nakamura the opportunity to contribute to this
special issue, T. Geszti, J. Hajdu, J. Kert\'esz, P. Pollner, P. Sz\'epfalusy, G. Tichy, 
G. Zar\'and and  A. Zawadovski  their interest and encouragement and
P. Dahlqvist the discussions. This work has been supported by the Hungarian Science Foundation 
OTKA (F019266/F17166/T17493), the Hungarian Ministry of Culture and 
Education (MKM 337) and the European Community under the PECO programme.


\begin{thebibliography}{99}
\bibitem{zero} The probability to hit a point like impurity is zero.
\bibitem{reg} The regularization removes the zero length trajectory
from the point to itself, which causes the divergence of the Green function
otherwise.
\end{thebibliography}
\end{document}